\documentclass[aps,prl,twocolumn,groupedaddress,showpacs]{revtex4}

\newcommand{\be} {\begin{equation}} 
\newcommand{\ee} {\end{equation}}
\newcommand{\bea} {\begin{eqnarray}}
\newcommand{\eea} {\end{eqnarray}}
\newcommand{\A} {{\mathcal{A}}}

\begin{document} 
\title{A Covariant Geometrical Representation of Quantum Interacting Electrons}
\author{Daniel C. Galehouse}
\email[]{dcg@uakron.edu} 
\affiliation{University of Akron, Akron, Ohio 44325}
\altaffiliation{15764 Galehouse Road, Doylestown, Ohio 44230}
\date{\today}
\begin{abstract}
A study of fundamental geometrical interactions shows that the Dirac
electron can be represented as a conformal wave.  A Riemannian space
is used, having coordinates that transform locally as spinors.  The
wave function becomes a gradient.  Anti-commuting matrices map the
eight-dimensional waves into five dimensions.  A projection into space
time gives the known gravitational and electromagnetic forces. The
electron transforms into a neutrino under hyper-rotations.  First
quantization is automatic. The theory is covariant and contains the
known electron interactions.
\end{abstract}
\pacs{2.40.Ky, 03.65.Pm, 11.10Kk, 12.15.-y} 
\maketitle

\section{Introduction \label{sec:intro}}

The objective of this article is to introduce a fundamental
geometrical description of the relativistic electron.  Following the
initial development by Dirac~\cite{dirac1,dirac2}, and subsequent
studies by mathematicians and physicists~\cite{badejehle}, the
understanding of fundamental spin has continued to be a major enigma
of modern physics.  The opportunity to describe the electron as a
Riemannian structure in a curvilinear space is made possible by the
development of a covariant five dimensional theory~\cite{fgqp} that
describes the interactions of quantum particles.  A recent
discussion~\cite{qgtis} gives a general overview and some mathematical
preliminaries.  The eight-dimensional formalism predicts the known
properties of electrons when projected through five dimensions onto
space-time.

The present work initiates the development of a class of exact
equations which describe the quantum properties of spin-1/2 particles.
The usual representation of point interactions of elementary particles
is to be replaced by geometrical transformations that follow the
particle field continuously through the interaction. To do this, an
individual Riemannian structure is assigned to each particle.  No
classical-like point objects are used.

Fields in five or eight dimensions are represented in a single
universal coordinate system.  From the curvature tensors, a first
quantized structure is generated without active
quantization~\cite{qfuft}, and a quantum field is developed without
reference to a classical basis.  This geometrical paradigm insures
compatibility with general relativity.  Mass appears during the
reduction from five to four dimensions.  It satisfies gravitation
equivalence and has mechanical inertia that is appropriate to quantum
particles.

An absolute equivalence for all interactions is assumed.  Internal
transformations convert between different types of force.  Each
particle has a full set of fields, including a metric, vector
potential, and a wave function.  These describe motion and
interaction.  The equivalence transformations are analogous with, and
in some cases identical to, the gauge transformations used in quantum
field theory.  The wave functions, and also the interaction
mechanisms, are associated with conformal factors that appear
naturally.  Characteristic linear wave equations come from the
invariant conformal waves.

Fundamental interactions are all taken to be time-symmetric.
Sufficient distant absorbing particles are assigned to account for the
retarded behavior of electromagnetism.  The resulting structure is a
set of mathematically determined, non-linear, interacting wave
fields. Second quantization is approached as a formalism to account
for multiple particles.  The combined set of fields evolves without
any assumption of fundamental statistics.  Photons and gravitons
acquire their discreet properties entirely from the countability of
the emitting particles.  No free fields are used.
  
\section{Five-Dimensional Theory\label{sec:fdt}}

Because some of the properties of five dimensional theory are required,
a short summary is given here.  The five coordinates, $x^m, m=0,\cdots
4$, can, in a local Riemannian system, be taken equal to
$(t,x,y,z,\tau)$.  The usual four-metric is 
\be 
d\tau^2 =g_{\mu
\nu}dx^\mu dx^\nu
\label{eq:propert}
\ee 
and becomes~\footnote{The sign convention is difficult for
historical reasons.  $\A_\mu$ is to be interpreted as a 
quantum-corrected kinetic four velocity.  
It must have the opposite sign as
the vector potential because the classical five metric is defined
using the vector potential $A_\mu$.}  \be 0 = g_{\mu \nu}dx^\mu
dx^\nu -(-\A_\mu dx^\mu - d\tau)^2 \equiv \gamma_{mn} dx^m dx^n
\label{eq:nully}
\ee 
which gives 
\be 
\gamma_{mn} \equiv \pmatrix{ g_{\mu \nu} - \A_\mu
\A_\nu & -\A_\mu \cr -\A_\nu & -1 \cr}
\label{eq:5metric}
\ee where, in this gauge, \be \A_\nu ={1 \over m} \left[{\partial
\over \partial x^\nu} \Im (\ln \psi) - e A_\nu \right]
\label{eq:velpot}
\ee

All curvature is taken to be conformally generated, so that the Ricci
tensor of $ \omega \gamma^{mn} $ for some function $\omega$ is
identically zero.  This implies that source terms exist and that they
can be calculated from $\omega$.  A particular choice gives the
gravitational and electrodynamic field equations. In four-dimensional 
form these are 
\bea 
R^{\alpha \beta} = 
 8 \pi \kappa
\bigglb( F^\alpha_{\phantom{\alpha}\mu}F^{\mu \beta} + m|\psi|^2 \A^\alpha
\A^\beta \nonumber \\
+ m|\psi|^2 { 1-\A^2 \over 2-\A^2} g^{\alpha \beta} \biggrb)
\label{eq:gft}
\eea
\be
F^{\beta \mu}|_\mu = 4 \pi e |\psi|^2 \A^\beta.
\label{eq:mft}
\ee 
The quantum field equation can be found by working from the
invariant curvature scalar.  Setting it to zero gives 
\bea 
{1 \over \sqrt{- \dot g}} 
\left(i\hbar {\partial \over \partial x^\mu} -eA_\mu \right) 
\sqrt { -\dot g} 
g^{\mu \nu } 
\left(i \hbar {\partial \over \partial x^\nu} -eA_\nu\right)\psi 
\nonumber  \\ 
= \left[ m^2 + {3 \over 16 }
\left( \dot R - {e^2 \over 4 m^2} F_{\alpha \beta} F^{\alpha \beta}\right) 
\right] \psi.
\label{eq:qft}
\eea 
Here, the fifth dimension has been reduced by setting derivatives
of the conformal factor with respect to the proper time equal to the
mass.  This reduction insures that identical particles have 
equal rest mass.
     
\section{Eight-Dimensional Derivation \label{sec:deriv}}

The Dirac equation must first be written in five-dimensional form.
The fifth anti-commuting spin matrix is assigned to the fifth
coordinate.  The commutation properties are
\bea \gamma^{mn}\delta_A^B = {1 \over 2} \{ \gamma^m , \gamma^n \}
 \equiv {1 \over 2} ( \gamma^{m \phantom{A} C} _{\phantom{m} A}
 \gamma^{n \phantom{C} B} _{\phantom{m} C} + \gamma^{n \phantom{A} C}
 _{\phantom{m} A} \gamma^{m \phantom{C} B} _{\phantom{m} C}) \nonumber \\
\hbox{for } m,n=0, \cdots, 4, \hbox{ and } A,B,C = 1, \cdots , 4.
\label{eq:anticom}
\eea 
The five-dimensional form of the Dirac equation is
used~\cite{qgtis}, 
\be \gamma^{m \phantom{A} B} _{\phantom{m} A} \,
{\partial \over \partial x^m} \Psi_B =0
\hskip .2 in \hbox{ or } \hskip .2 in \gamma^m {\partial \over
\partial x^m} \Psi =0 .
\label{eq:deq}
\ee 
To relate this to properties of conformal transformations in eight
space, we choose eight real coordinates $ \xi_r^A , \xi_i^A $ for $ A
= 1 \cdots 4 $, and consider the associated conformally flat space.
Let the conformal factor be denoted by $\omega$.  The condition that
the scalar curvature is zero constitutes a second-order differential
equation for $\omega$ and provides a Riemannian characterization of
the conformal waves.  The equation can be written in modern spinor
notation.  The eight coordinates must be arranged in complex pairs.
\be 
\xi^A = \xi^A_r + i\xi^A_i, \xi^{ \bar
A} = \xi^A_r - i\xi^A_i ,\hskip .2 in A=1 \cdots 4
\label{eq:cpairs}
\ee 
where the bar of conjugation is applied to the index for consistency
with existing conventions~\cite{vdwaerd}.  The fundamental form is
$d \xi^A d \xi^{ \bar B } \epsilon_{A \bar B}$
with spinor metric \be \epsilon_{A \bar B}=\epsilon^{A \bar B}=
\hbox{diag}(1,1,-1,-1).
\label{eq:smetric}
\ee
The curvature equation is 
\be 
\epsilon^{ \bar
AB} {\partial \over \partial \xi^{ \bar A }} {\partial \over \partial
\xi^B} \Psi =0
\label{eq:scurve} 
\ee 
where $ \omega = \Psi^p$ with $p = 4/(n-2) = 2/3$.

Locally, the coordinates in the neighborhood of a point in five-space
are related to a corresponding point in eight-space by the
differential relation 
\be 
dx^m = \zeta^{A} 
\gamma^{m \phantom{A} B}_{\phantom{m} A} 
d \xi^{ \bar C} \epsilon_{ \bar CB}+ 
d \xi^{A}\bar \gamma^{m B} _{\phantom{m B} A} 
\zeta^{ \bar C} \epsilon_{ \bar CB}
\label{eq:8to5} 
\ee 
Because the eight-dimensional displacements are summed with
conjugates, the five space displacements are always real.  Furthermore
Lorentz rotations, extended to five dimensions, are mapped in the
usual way from the spinor space of displacements $d\xi^A$, $d
\xi^{\bar A}$ to $dx^m$.  It is the simplest transformation law which
satisfies these conditions.  The parameter $\zeta$ specifies the
relative orientation of the spin space.  It is equivalent to the use
of a four- or five-dimensional spin frame.

The quantity 
\be 
\Psi_B = { \partial \Psi \over \partial \xi^B }
\label{eq:dspinor}
\ee 
is taken as the Dirac spinor wave function which, following
equation~(\ref{eq:scurve}), satisfies the first order equation 
\be
\epsilon^{\bar AB} { \partial \Psi_B \over \partial \xi^{ \bar A}} =0
\label{eq:8ddeq}
\ee 
Using the chain rule 
\[ 
\displaystyle {\partial \over \partial \xi^{
\bar A}} = {\partial x^m \over \partial \xi^{ \bar A} } {\partial
\over \partial x^m} 
\] 
from the coordinate transformation
(\ref{eq:8to5}), this becomes \be \epsilon^{ \bar AB} \gamma^{m
\phantom{A} C} _{\phantom{m} D} \zeta^D \, \epsilon_{\bar AC}{\partial
\Psi_B \over \partial x^m} \equiv \zeta^D \left( \gamma^{m \phantom{D}
B}_{\phantom{m} D} \, {\partial \Psi_B \over \partial x^m} \right) = 0
\label{eq:gdeq}
\ee 
The complex conjugates $\xi^{\bar A}$ are treated as independent
of the $\xi^A$'s during differentiation.  The quantity in parenthesis
is identified with equation (\ref{eq:deq}), and is interpreted as
characterizing the eight-dimensional space in an orientation
independent way. Equation~(\ref{eq:gdeq}) should be satisfied for any
value of $\zeta$.

Using equation (\ref{eq:dspinor}), a local plane wave solution 
of (\ref{eq:qft}) can be converted to a
solution of (\ref{eq:deq}) by differentiation.  
\be 
\Psi = e^{i(\omega t - \roarrow k \cdot \roarrow x -  m \tau)} 
\equiv e^{ik_m x^m}, \hskip .2 in k_m=(\omega , \roarrow k ,  m)
\label{eq:kgpw}
\ee 
\be 
\Psi_A \equiv {\partial \Psi \over \partial \xi^A} = \Psi i
k_m {\partial x^m \over \partial \xi^A}  
= i \Psi (k_m \gamma^{\dagger m})\zeta^\dagger
\label{eq:dsol}
\ee 
This is the form for a free electron or positron with arbitrary spin 
when written in the basis implied by the matrices of 
equation (\ref{eq:deq}).  In this way, it is assured that the particle
is locally a Dirac electron.

This derivation involves a minimum of physical assumptions and has 
no explicitly performed first quantization.  
Gravitational and electromagnetic interactions are included when the 
$\gamma^{m \phantom{A} B}_{\phantom{m} A}$ 
matrices are defined.  Moreover, the conformal variations that
generate external source terms in the five-dimensional theory imply,
through (\ref{eq:8to5}), that they are equivalent to analogous
conformal transformations in the eight-dimensional space.  Forces
other than electromagnetism or gravitation must be present because the
matrix of second order coordinate derivatives has become larger.  Weak
interactions are expected because they are known to be described with
the Dirac theory, and equivalence requires that such additional forces
be gauged to the others.
  
In local Cartesian coordinates, and using standard notation for the
$\gamma$'s, the six quantities 
\bea 
A^\mu = \psi^\dagger \gamma^\mu \psi, \hbox{
for } \mu = 0,1,2,3, \nonumber \\
\hbox{ with } A^4 =i\psi^\dagger \gamma^5 \psi,
\hbox{ and } A^5 =i\psi^\dagger \psi
\label{eq:pkterms}
\eea 
following~\cite{oveblen,vfock} combine into a quadratic invariant
which, in modern notation, is 
\be
(A^0)^2-(A^1)^2-(A^2)^2-(A^3)^2=(A^4)^2+(A^5)^2.
\label{eq:pki}
\ee 
Term by term, in the classical limit, it corresponds to the
relation $ E^2 - p^2 =m^2 $.  Two of the six quantities must make up
the mass.  If these two quantities cancel, a zero-mass particle
results.  The condition is 
\bea 
0 = (A^4)^2 +(A^5)^2 \equiv (A^5
-iA^4)(A^5 + iA^4) \nonumber \\
\equiv [\psi^\dagger ( 1 + \gamma^5) \psi] \cdot
[\psi^\dagger ( 1 - \gamma^5) \psi]
\label{eq:zmass}
\eea 
One of the factors must be zero, giving either a neutrino or an
anti-neutrino.  Equation (\ref{eq:8ddeq}) has a zero-mass solution,
and it must transform into the electron solution under a
hyper-rotation.  This inter-transformation is identified with weak
isospin.  Because (\ref{eq:qft}) and (\ref{eq:deq}) can only apply to
a system of particles of fixed charge-to-mass ratio, neutrinos and
anti-neutrinos must have, respectively, the same charge-to-mass ratio
as the electron and positron. A more detailed representation of the
weak force as a conformal effect will be discussed at a later time.

\section{Interpretation\label{sec:interp}}

Some of the properties of the postulated eight-dimensional physical
space are not yet understood.  The dimensional reduction does not
follow the common Kaluza-Klein paradigm since spin properties persist
from the base space. The success of the reduction depends on starting
with a sufficiently simple structure in eight dimensions. If both the
five- and eight-dimensional spaces are conformally flat, integrability
issues are minimized.  The assumption of conformal flatness appears to
be rich enough to describe gravitational, electromagnetic, and weak
interactions, but greater complexity may be possible.  It is not know
what additional phenomenological predictions can be expected.

The proposed continuing methodology is to write strong interactions in
eight-dimensional form, assume a universal geometry of isospin, and
then try to understand what additional representations are needed.
Particle transmutation by isospin hyper-rotation is distinct from the
compositional change seen in bound systems, and implies a general
theory of mass.  Historically, each increment in the number of
dimensions has demonstrated an increasingly richer collection of
physical observations.  The increase from five to eight dimensions may
be enough to include the quarks or the gluonic interactions.  More
sophisticated internal representations of particle mass are expected,
especially if hadrons or quarks are involved.  Specific values may
appear with particle properties to match.

The additive terms to the square of the electron mass, as seen in
(\ref{eq:qft}), are characteristic of higher-dimensional projected
theories.  The coefficient depends on the number of dimensions.  The
terms cannot be eliminated if, in a specific dimensionality, all
coordinates are formally equivalent.  This implies a propagational
mass that is distinct from the rest mass.  In a strong gravitational
or electromagnetic field, the effective mass that appears in the local
equation of motion is different from that in the absence of the
interacting field.  This effect is unobservably small for the
electron, but may be measurable, eventually, for neutrinos. Possible
contributions from the flavor interactions are not understood, and may
be important.  Also, even though strong interactions are not usually
associated with electrons, they may contribute as well.

Fiber bundles, originally developed to describe spin in a curvilinear
setting, continue to be relevant.  An implication of this construction
is that a bundle may be symptomatic of the residual effects of
dimensional reduction.  Some Yang-Mills theories may fall into this
class.  Both mathematical and physical issues are involved. It is
usually claimed that the graviton, because it is spin two, cannot be
the Higgs particle.  Here the graviton is grouped with the photon to
make a five-scalar. The usual condition may be overly restrictive.
The Higgs mechanism should connect with the mass mechanisms of quantum
theory and general relativity.  A curvilinear methodology is required,
and may naturally bring the ideas together.
 
\section{Conclusion\label{sec:concl}}

The Dirac electron theory is reformulated in an eight-dimensional
coordinate space.  The eight parameters combine into four complex
pairs that transform locally according to spinor representations of
the extended Lorentz group. Anti-commuting Dirac matrices are used to
map local complex displacements onto five-dimensional space.  Each
particle has a separate set of interacting gravitational and
electromagnetic fields.  The wave function is represented by the
eight-gradient of a conformal parameter.  If this parameter satisfies
the condition of zero scalar curvature, the Dirac equation appears
from the geometry. Neutrino solutions satisfy the same curvature
condition and transform into electrons under hyper-rotation. This gives
the isospin-changing weak interactions an explicit mechanical model.
The eight dimensional formalism may be suitable for strong
interactions.  This construction augments the known explicit
geometrical descriptions of quantum electrons moving in gravitational
and electromagnetic fields to include the weak forces. Until such time
as strong interactions might be observed, this provides a complete
description of the electron.

\end{document}